\documentclass[aps,prapplied,longbibliography,superscriptaddress,floatfix,twocolumn]{revtex4-1}

\usepackage{graphicx}
\usepackage{latexsym}
\usepackage{graphicx}
\usepackage{times}
\usepackage{color}
\usepackage{amsmath}
\usepackage{dcolumn}
\usepackage{mathbbol}
\usepackage{latexsym,amsmath,amssymb,bm,euscript}
\usepackage{xr}
\usepackage{lipsum}
\usepackage[normalem]{ulem}
\usepackage{relsize}
\usepackage{hyperref}
\hypersetup{
    colorlinks,%
    citecolor=blue,%
    filecolor=blue,%
    linkcolor=blue,%
    urlcolor=blue
}

\newcommand{\da}{\dagger}
\newcommand{\an}[1]{\hat{a}_\text{#1}}
\newcommand{\ad}[1]{\hat{a}_\text{#1}^\dagger}
\newcommand{\iqp}{\hat{I}_\text{qp}}
\newcommand{\re}{\text{Re}}
\newcommand{\im}{\text{Im}}
\newcommand{\dw}{\Delta/\hbar\omega_0}

\begin{document} 
\title{Parametric amplification and squeezing with an ac- and dc-voltage biased superconducting junction}
\author{Udson C. Mendes}
\email{udsonmendes@ufg.br}
\affiliation{Institut quantique and D\'{e}partment de Physique, Universit\'{e} de Sherbrooke, Sherbrooke, Qu\'{e}bec J1K 2R1, Canada}
\affiliation{Service de Physique de l'Etat Condens\'e, CEA, CNRS, Universit\'e Paris-Saclay, CEA Saclay, 91191 Gif-sur-Yvette, France}
\author{S\'{e}bastien Jezouin}
\affiliation{Institut quantique and D\'{e}partment de Physique, Universit\'{e} de Sherbrooke, Sherbrooke, Qu\'{e}bec J1K 2R1, Canada}
\author{Philippe Joyez}
\affiliation{Service de Physique de l'Etat Condens\'e, CEA, CNRS, Universit\'e Paris-Saclay, CEA Saclay, 91191 Gif-sur-Yvette, France}
\author{Bertrand Reulet}
\affiliation{Institut quantique and D\'{e}partment de Physique, Universit\'{e} de Sherbrooke, Sherbrooke, Qu\'{e}bec J1K 2R1, Canada}
\author{Alexandre Blais}
\affiliation{Institut quantique and D\'{e}partment de Physique, Universit\'{e} de Sherbrooke, Sherbrooke, Qu\'{e}bec J1K 2R1, Canada}
\affiliation{Canadian Institute for Advanced Research, Toronto, Canada}
\author{\\  Fabien Portier}
\affiliation{Service de Physique de l'Etat Condens\'e, CEA, CNRS, Universit\'e Paris-Saclay, CEA Saclay, 91191 Gif-sur-Yvette, France}
\author{Christophe Mora}
\affiliation{Laboratoire Pierre Aigrain, \'Ecole normale sup\'erieure, PSL Research University, CNRS, Universit\'e Pierre et Marie Curie, Sorbonne
Universit\'es, Universit\'e Paris Diderot, Sorbonne Paris-Cit\'e, 24 rue Lhomond, 75231 Paris Cedex 05, France}
\author{Carles Altimiras}
\affiliation{Service de Physique de l'Etat Condens\'e, CEA, CNRS, Universit\'e Paris-Saclay, CEA Saclay, 91191 Gif-sur-Yvette, France}
\date{\today}

\begin{abstract}
We theoretically investigate a near-quantum-limited parametric amplifier based on the nonlinear dynamics of quasiparticles flowing through a superconducting-insulator-superconducting junction. Photon-assisted tunneling, resulting from the combination of dc- and ac-voltage bias, gives rise to a strong parametric interaction for the electromagnetic modes reflected by the junction coupled to a transmission line. We show phase-sensitive and phase-preserving amplification, together with single- and two-mode squeezing. For an aluminum junction pumped at twice the center frequency, $\omega_0/2\pi=6$~GHz, we predict narrow-band phase-sensitive amplification of microwaves signals to more than 20 dB, and broadband phase-preserving amplification of 20 dB over a 1.2 GHz 3-dB bandwidth. We also predict single- and two-mode squeezing reaching more than -12 dB over 5.3 GHz 3-dB bandwidth. Moreover, with a simple impedance matching circuit, we demonstrate 3 dB bandwidth reaching 4.3 GHz for 20 dB of gain. A key feature of the device is that its performance can be controlled in-situ with the applied dc- and ac-voltage biases.
\end{abstract}

\maketitle

\section{Introduction}

Many of the advances of quantum computation based on superconducting qubits rely on the ability to readout the qubit state by measuring microwave photons leaking out of a superconducting resonator \cite{wallraff-schoelkopf-nature2004}. Thanks to the development of near-quantum-limited Josephson parametric amplifiers (JPAs) \cite{beltran-lehnert-apl2007,bergeal-devoret-nature2010,zhou-esteve-prb2014,eichler-wallraff-prl2014,jebari-hofheinz-arxiv2017}, high-fidelity single-shot qubit readout is now possible \cite{walter-wallraff-prappl2017,vijay-siddiqi-prl2011}. These amplifiers are, moreover, finding use in a wide range of applications, from measuring quantum features in the radiation emitted by mesoscopic conductors \cite{zakka-portier-prl2010,gasse-reulet-prl2013,stehlik-petta-prappl2015,westig-portier-prl2017,simoneau-reulet-prb2017}, to the detection of small ensembles of electronic spins \cite{bienfait-bertet-prx2017}, and even to the search for dark matter \cite{brubaker-carosi-prl2017}. JPAs are also versatile sources of single- and two-mode squeezed states \cite{beltran-lehnert-nphys2008,eichler-wallraff-prl2014}, which have been used to confirm decade old predictions in quantum optics \cite{murch-siddiqui-nature2013,toyli-siddiqi-prx2016}, and to improve electron-spin resonance spectroscopy \cite{bienfait-bertet-prx2017}. Theoretically, squeezed states were proposed as a resource to improve qubit readout and to perform high-fidelity gates \cite{didier-blais-clerk-prl2015,puri-blais-prl2016,royer-blais-quantum2017}, or as  basis for continuous variable quantum computing \cite{braunstein-loock-rmp2005,grimsmo-blais-npjqi2017}.

Current JPAs are able to amplify signals to more than 20 dB, and to squeeze vacuum fluctuations by $7$ dB ($12$ dB) in single- (two-) mode experiments \cite{boutin-blais-prappl2017,eichler-wallraff-prl2014}. However, in these devices, the amplification bandwidth is limited to hundreds of megahertz \cite{westig-klapwijk-prapplied2018,mutus-martinis-apl2014,roy-vijay-apl2015}. At the price of increasing device fabrication complexity, much larger amplification bandwidth, over $\sim 3$ GHz, has been demonstrated with the recently developed Josephson traveling wave parametric amplifier \cite{macklin-siddiqi-science2015}. The development of a simpler quantum-limited microwave amplifier, generating far-separated two-mode squeezed states and capable of amplifying signals over gigahertz bandwidths, is still needed to further advance quantum information processing science. It would also be an important tool to better characterize the radiation emitted by mesoscopic conductors, for which there is an increasing body of interesting predictions \cite{beenakker-schomerus-prl2004,armour-rimberg-prl2013,leppakangas-johansson-prl2015,mendes-mora-njp2015,mendes-mora-prb2016}. Here, we propose a simple broadband parametric amplifier, consisting of a single dc- and ac-voltage biased superconductor-insulator-superconductor (SIS) junction. The device can be operated in both phase-sensitive and phase-preserving modes and can be used for near-quantum-limited amplification and two-mode squeezing in few GHz bandwidth.

\begin{figure}[t]
\begin{center}
\includegraphics[width=0.4\textwidth]{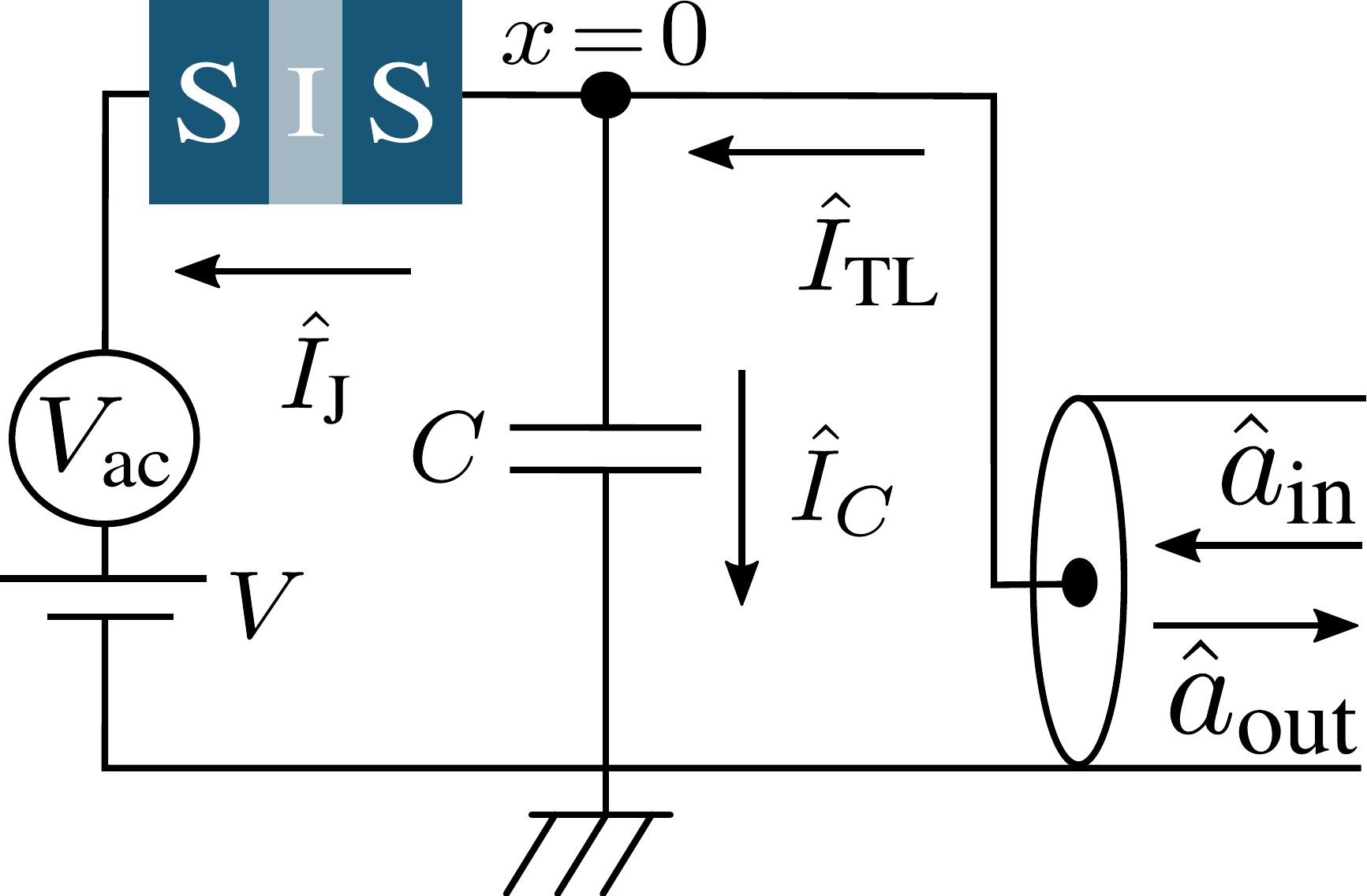}
\caption{Electrical scheme of the device: a SIS junction is dc-voltage biased close to the onset of quasiparticle transport $eV_\text{dc} \sim 2\Delta$, while it is ac pumped with a single tone $V_\text{ac} \cos(2\omega_0 t)$. Current conservation at the coupling node ($x=0$) allows to relate the transmission line outgoing field $a_\text{out}$ with its incoming field $a_\text{in}$ and the current flowing through the junction $\hat{I}_\text{J}$. \label{fig1a}}
\end{center}
\end{figure}

The proposed setup, illustrated in Fig.~\ref{fig1a}, operates as an amplifier in reflection mode. Parametric amplification is possible by taking advantage of the strong nonlinearity of the transport characteristics of the junction. To this end, we consider a dc-voltage bias $V_\text{dc}$ smaller than twice the superconducting energy gap $\Delta$. At this bias point, the junction behaves as an open circuit and the conduction of quasiparticles is enabled by pumping it with a sinusoidal ac-voltage $V_\text{ac}(t) =V_\text{ac}\cos(2\omega_0 t)$, where $\omega_0$ is the center measurement frequency. This voltage combination gives rise to modulations of the admittance of the junction $Y_n(\omega-2n\omega_0)$, with $n$ the $n$-th sideband of the pump, which generates ac-quasiparticle current at frequency $\omega-2n\omega_0$. As we will show, the most important terms are: first, $Y_1(\omega - 2\omega_0)$ which is related to the coherent conversion process of one quanta of energy $2\hbar\omega_0$ from the pump to two photons of frequencies $\omega \approx\omega_0$, and second $Y_0(\omega)$ which is related to single-photon coherent (imaginary part) and dissipative (real part) response due to the tunneling of quasiparticles. By appropriately choosing the dc-voltage below twice the superconducting gap, we demonstrate that it is possible to make $|Y_1(\omega-2\omega_0)|$ large enough while keeping Re$Y_0(\omega)$ close to zero in order to generate parametric amplification with near quantum-limited noise and squeezing. Furthermore, Im$Y_0(\omega)$ gives rise to a large frequency-dependent impedance mismatch which limits the gain and squeezing bandwidths. Fortunately, as  we will present in Sec.~\ref{impeng}, this frequency-dependent impedance mismatch can be dealt with an impedance matching scheme.

It is surprising that even though SIS junctions are routinely used as high-frequency microwave quantum-limited mixers \cite{zmuidzinas-richards-PIEEE2004,westig-honingh-sst2011}, exploiting a very similar principle, their operation as parametric amplifiers have been mostly disregarded \cite{lee-apl1982,devyatov-zorin-jap1986}. Here, we use the input-output formalism \cite{yurke-denker-pra1984}, together with photon-assisted tunneling theory \cite{tien1963} to compute the parametric amplification and squeezing properties of an ac- and dc-voltage biased SIS junction \cite{trucker-feldman-rmp1985}. The resulting Heisenberg-Langevin equations \cite{gardiner-collet-pra1985} are numerically solved, allowing us to explore parametric amplification far from the small detuning limit considered previously \cite{lee-apl1982,devyatov-zorin-jap1986}. For an aluminum junction ($\Delta = 180~\mu\text{eV} \sim h \times 43.5$ GHz) pumped at $2\omega_0 =2\pi \times 12$ GHz and operated at temperatures $T \ll \Delta/k_B$, with $k_B$ the Boltzmann constant, we find that when operated in experimentally relevant conditions the device can produce more than 20 dB of phase-preserving and phase-sensitive gain, and $\sim 13$~dB of single- and two-mode squeezing. In the phase-preserving mode, the 3 dB gain bandwidth exceeds 1.2 GHz and, therefore, it is twice as large as the bandwidth of the broadband impedance engineered JPA \cite{roy-vijay-apl2015}. Moreover, the 3 dB gain and squeezing bandwidths can be increased to 4.3 GHz with impedance matching schemes to compensate both the geometrical capacitance $C$ and dynamical susceptance Im$Y_0$ of the junction \cite{roy-vijay-apl2015}.
\begin{figure}[t]
\begin{center}
\includegraphics[width=0.4\textwidth]{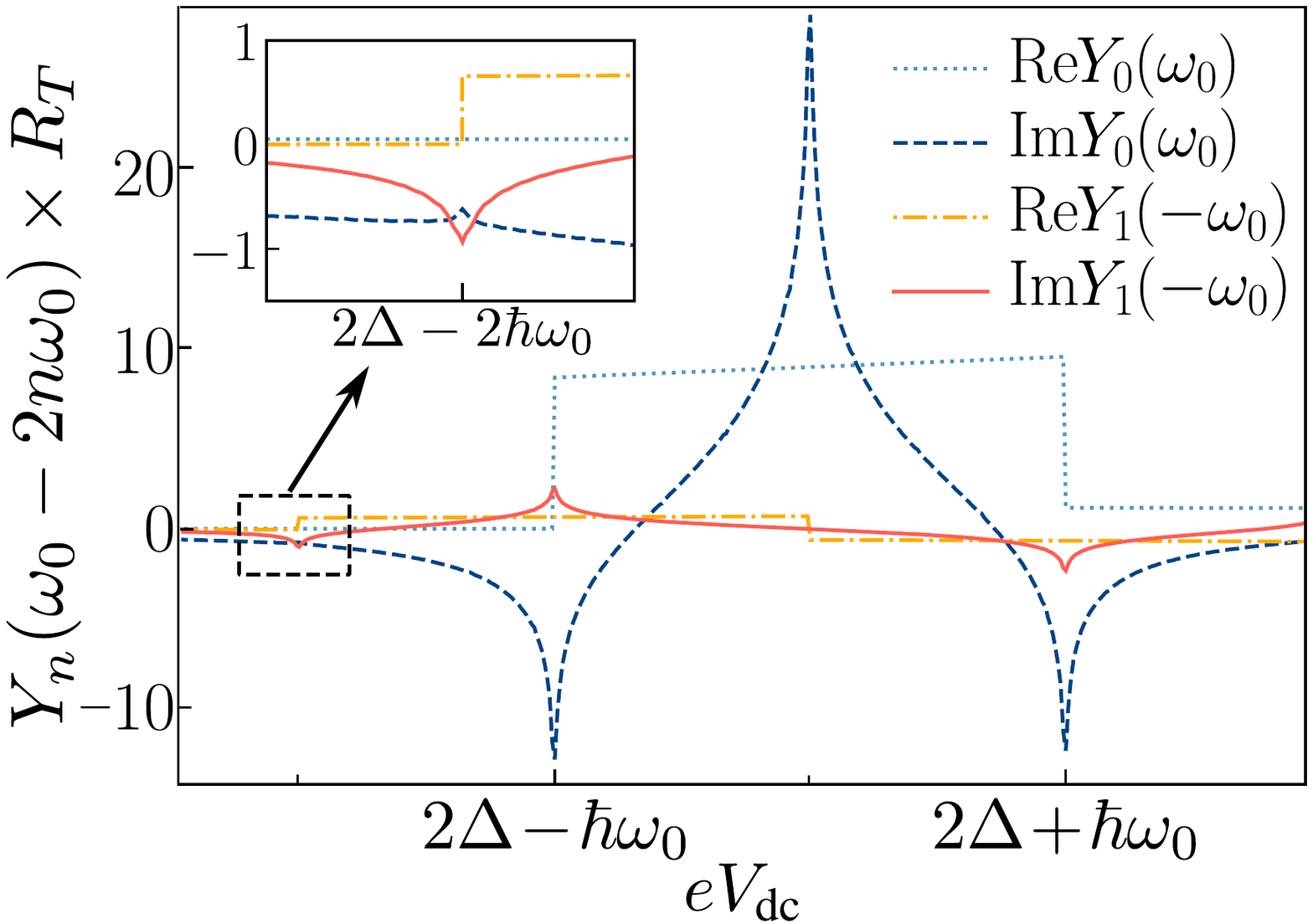}
\caption{Nonlinear admittance as a function of the dc voltage for an aluminum junction ($\Delta = 180~\mu\text{eV} \sim h \times 43.5$ GHz), $eV_\text{ac} = 0.155 \times 2\hbar\omega_0$ and $\omega = \omega_0 = 2\pi \times 6$ GHz. On the one hand, as shown in Sec.~\ref{model}, Im$Y_0(\omega_0)$ and Re$Y_0(\omega_0)$ give rise respectively to coherent and dissipative absorption and emission of single photons. However, the pumping gives rise to a parametric interaction characterized by a sizable nonlinear admittance $Y_1(-\omega_0)$. For dc-voltages $eV_\text{dc} \sim 2(\Delta-\hbar\omega_0)$ and with the appropriate ac-voltage amplitude, the nonlinear susceptance $Y_1(-\omega_0)$ dominates over the single-photon coherent and dissipative response $Y_0(\omega)$ giving rise to parametric amplification and squeezing (see inset). \label{fig1b}}
\end{center}
\end{figure}

This article is organized as follows: In Sec.~\ref{model} we describe the input-output formalism used to characterize the device. Section~\ref{resultsA} presents results for an ideal SIS junction for which the transport response rises steeply at $eV_\text{dc}=2\Delta-n\hbar\omega_0$. The effects of the low-frequency noise, which captures most non-ideal effects, on the amplifier is described in Sec.~\ref{resultsB}. An approach to improve further the performances of the amplifier relying on impedance engineering is presented in Sec.~\ref{impeng}. Final remarks are presented in Sec.~\ref{Final remarks}.

\section{Model \label{model}}
We consider a SIS junction in parallel with its capacitance and connected to a transmission line (TL), see Fig.~\ref{fig1a}. The total Hamiltonian of the device is $\hat{H} = \hat{H}_\text{qp} + \hat{H}_\text{ee} + \hat{H}_\text{t}$, with
\begin{equation} \label{Hqp}
 \hat{H}_\text{qp} = \sum_l \epsilon_l \hat{c}_l^\dagger \hat{c}_l + \sum_r \epsilon_r \hat{c}_r^\dagger \hat{c}_r 
\end{equation}
describes the quasiparticles in the left ($l$) and right ($r$) superconductors forming the junction. In this expression, $\hat{c}_{l(r)}$ annihilates a quasiparticle of energy $\epsilon_{l(r)}$ in the left (right) superconductor. The above Hamiltonian describes the dynamics of quasiparticles. However, Cooper pairs are also present and interact with the electromagnetic field. Since we are interested in signals of frequency $\omega_0 \ll \Delta/\hbar$ amplified or squeezed by operating the junction close to the onset of quasi-particle transport $eV_\text{dc} \approx 2\Delta$, we neglect the effects from the tunneling of Cooper pairs whose Josephson frequency is $\sim 4\Delta/\hbar \gg 2\omega_0$. Experimentally, the effects of Cooper pairs can be further suppressed by passing one flux quantum within a SIS junction \cite{trucker-feldman-rmp1985}. This suppression can be done either fabricating the junction in a SQUID geometry or by applying an in-plane magnetic field to junction \cite{westig-honingh-sst2011}.

The Hamiltonian of the electromagnetic environment, which includes the TL and the capacitor $C$ of the junction, is 
\begin{equation} \label{Hee}
 \hat{H}_\text{ee} = \frac{\hat{Q}^2(x=0)}{2C} + \int_0^\infty dx \left[ \frac{1}{2L_0} \left(\frac{\partial \hat{\Phi}(x)}{\partial x}\right)^2 + \frac{\hat{Q}^2(x)}{2C_0} \right].
\end{equation}
The first and second terms account for the charging energy of the capacitor and the TL dynamics. The charge $\hat{Q}(x)$ and flux $\hat{\Phi}(x)$ operators obey $[\hat{\Phi}(x)$,$\hat{Q}(x^\prime)] = i\hbar\delta(x-x^\prime)$. The TL characteristic impedance $Z_0 = \sqrt{L_0/C_0}$ is defined in terms of its inductance $L_0$ and $C_0$ capacitance per unity of length. Finally, the last term of total Hamiltonian is the tunneling of quasiparticles dressed by the environment and it takes the form \cite{devoret-urbina-prl1990}
\begin{equation} \label{tunnel}
\hat{H}_\text{t} = \hat{\mathcal{T}}(t) \exp[i e\hat{\Phi}(t)/\hbar] + \text{h.c.},
\end{equation}
with $\hat{\mathcal{T}}(t) = \sum_{l,r}t_{lr}\hat{c}_l^\dagger\hat{c}_r \exp \{i [eV_\text{dc} t/\hbar + \varphi_{ac}(t)]\}$ the tunneling operator transferring a quasiparticle from the right to the left side of the junction with tunneling probability amplitude $t_{lr} \equiv t_{rl}$. The phase $\varphi_\text{ac}(t) = (eV_\text{ac}/2\hbar\omega_0)\sin(2\omega_0 t)$ takes into account the ac-voltage and $\hat{\Phi}(t) \equiv \hat{\Phi}(t,x=0)$ is the TL flux at the position of the junction . 

In the interaction picture with respect to $\hat{H}_\text{t}$, this TL flux is written in terms of the incoming $\an{in} [\omega]$ and outgoing $\an{out} [\omega]$ fields as \cite{yurke-denker-pra1984}
\begin{align} \label{phase}
\hat{\Phi}(t,x) &= \sqrt{\frac{\hbar Z_0}{4\pi}} \int_{0}^{\infty} \frac{d\omega}{\sqrt{\omega}} \left[ \an{in}[\omega]  e^{-i(\omega t +k_\omega x) } \right.  \nonumber \\
&\left.  + \an{out}[\omega]  e^{-i(\omega t- k_\omega x)}  + \text{ h.c.} \right],
\end{align} 
with $k_\omega = \omega \sqrt{L_0 C_0}$ the wave number. The incoming (outgoing) field obeys the commutation relation $[\an{in (out)}[\omega],\ad{in (out)}[\omega^\prime]] = \delta(\omega - \omega^\prime)$.

To characterize the radiation emitted by the junction, we use the input-output formalism adapted to circuits coupled to quantum conductors \cite{leppakangas-johansson-prl2015,grimsmo-blais-prl2106,mora-portier-prb2017}. To obtain the input-output boundary condition, the first step is to derive the Heisenberg equations of motion for both TL charge and flux at the position of the junction ($x=0$). The resulting equations are combined to express current conservation
\begin{equation} \label{bond0}
C \ddot{\hat{\Phi}}(t,x=0) - \frac{1}{L_0} \frac{\partial \hat{\Phi}(t,x)}{\partial x} \bigg|_{x=0} = \hat{I}_\text{J}(t),
\end{equation}
which connects the outgoing field to the incoming field and the current of the junction $\hat{I}_\text{J}(t)$. Using Eq.~\eqref{phase}, the above equation can be expressed as an input-output relation
\begin{equation} \label{inout0}
 \an {out}[\omega] = \frac{1 + i Z_0 C\omega}{1 - i Z_0 C\omega} \an {in}[\omega] + i \sqrt{\frac{Z_0}{\pi \hbar \omega}} \frac{\hat{I}_\text{J}[\omega]}{1 - i Z_0 C\omega}, 
\end{equation}
where $\hat{I}_\text{J}[\omega]$ is the Fourier transform of $\hat{I}_\text{J}(t)$. The first term describes the phase shift due to the reflection of the input field by the capacitor. The last term characterizes the radiation emitted by the junction and it gives rise to both current fluctuations and a deterministic response to the applied voltages \cite{grimsmo-blais-prl2106,mora-portier-prb2017}.

The current $\hat{I}_\text{J}[\omega]$ depends not only on the dc- and ac-voltages, but also on the TL voltage, $\hat{V}_\text{TL}(t) = \dot{\hat{\Phi}}(t)$, implying that $\hat{I}_\text{J}[\omega]$ and $\an {in}[\omega]$ do not commute. To circumvent the non-commutation between $\hat{I}_\text{J}[\omega]$ and $\an {in}[\omega]$, we take advantage of the weak TL-junction coupling $\sqrt{\pi Z_0/R_K} \ll 1$ (for a typical $Z_0=50~\Omega$ TL impedance \cite{zakka-portier-prl2010,gasse-reulet-prl2013}, with $R_K \simeq 25.8$ k$\Omega$ the quantum of resistance) and compute $\hat{I}_\text{J}[\omega]$ to second-order in the TL-junction coupling \cite{mendes-mora-njp2015,grimsmo-blais-prl2106}. In the time domain and for weak coupling to the environment, the current operator, in the Heisenberg picture, takes the form
\begin{equation} \label{current0}
\hat{I}_\text{J}(t) =\hat{I}_\text{qp}(t) + \frac{e^2}{\hbar^2}\hat{H}_\text{qp}^\text{t}(t)\hat{\Phi}(t),  
\end{equation}
where $\hat{I}_\text{qp}(t) = i(e/\hbar) (\hat{\mathcal{T}}^\dagger(t) - \text{h.c.})$ and $\hat{H}_\text{qp}^\text{t}(t) = \hat{\mathcal{T}}(t) + \text{h.c.}$ are respectively the quasiparticle current and tunneling operators in the absence of TL voltage. The final step is to time-evolve $\hat{I}_\text{J}(t)$. Using linear response theory, the quasiparticles operators are time evolved, in the interaction picture, with the interaction Hamiltonian $H_\text{int}(t)= -\hat{I}_\text{qp}^{i}(t)\hat{\Phi}^{i}(t)$, with superscript $i$ meaning the operator in interaction picture. Once more, we take advantage of the weak TL-junction coupling and the short interaction time between the quasiparticles and photons to expand the time-evolution operator to first-order in the $\hat{\Phi}^{i}(t)$ to obtain 
\begin{align} \label{curr1}
\hat{I}_\text{J}(t) &= \hat{I}_\text{qp}^i(t) + \frac{e^2}{\hbar^2}\hat{H}_\text{qp}^{t(i)}(t)\hat{\Phi}^i(t) \nonumber \\
&-\frac{i}{\hbar}\int_{-\infty}^{t}\hat{\Phi}^i(t^\prime)[ \hat{I}_\text{qp}^i(t^\prime), \hat{I}_\text{qp}^i(t)]dt^\prime.
\end{align}
The first term is the current operator due to quasiparticle tunneling and it depends on the ac- and dc-voltage biases, while the two last terms describe modifications of the current fluctuations due to TL field. In the weak interaction limit considered here, we average over the quasiparticle operators in the terms proportional to $\hat{\Phi}^i(t)$, i.e., 
\begin{align} \label{curr2}
\hat{I}_\text{J}(t) &= \hat{I}_\text{qp}^i(t) + \frac{e^2}{\hbar^2}\langle \hat{H}_\text{qp}^{t(i)}(t) \rangle \hat{\Phi}^i(t) \nonumber \\
&-\frac{i}{\hbar}\int_{-\infty}^{t}\hat{\Phi}^i(t^\prime)\langle[ \hat{I}_\text{qp}^i(t^\prime), \hat{I}_\text{qp}^i(t)]\rangle dt^\prime.
\end{align}
This approximation is equivalent to the Born-Markov approximation \cite{mendes-mora-njp2015,grimsmo-blais-prl2106}. Under these approximations, the current operator is written in frequency space as
\begin{equation} \label{current}
\hat{I}_\text{J}[\omega] = \hat{I}_\text{qp}^i[\omega] - \sum_{n=-\infty}^\infty Y_{n}(\omega- 2n \omega_0) \hat{V}_\text{TL}^i[\omega - 2n\omega_0],
\end{equation}
with $\hat{I}_\text{qp}^i[\omega]$ and $\hat{V}_\text{TL}^i[\omega]$ the Fourier transforms \footnote{Fourier transform of the operator $\hat{I}[t]$ is defined as $\hat{I}[\omega] = \int_{-\infty}^\infty \hat{I}(t) e^{i\omega t} dt$, in accordance with quantum mechanics definition.} of $\hat{I}_\text{qp}^i(t)$ and $\hat{V}_\text{TL}^i(t)$, respectively. The generalized admittance 
\begin{align} \label{admittance}
Y_n(\omega) &= \frac{i}{\hbar} \int \frac{d\omega_1}{2\pi}\frac{S_n(\omega_1)-S_n(2n\omega_0-\omega_1)}{(\omega_1 +\omega + i0^+)(\omega_1 + i0^+)},
\end{align}
relates the current response at frequency $\omega$ to the TL-voltage dynamics at frequency $\omega - 2n\omega_0$ \cite{mora-portier-prb2017}. The admittance is defined in terms of the photon-assisted current-current correlator $\langle \hat{I}_\text{qp}^i(\omega) \hat{I}_\text{qp}^i(\omega^\prime) \rangle = 2\pi \sum_n S_n(\omega) \delta(\omega+\omega^\prime - 2n\omega_0)$ where 
\begin{align} \label{noise-PE}
S_n(\omega) & =\frac{1}{2} \sum_{n_1=-\infty}^{\infty} J_{n_1}(\rho) [J_{n_1+n}(\rho) S_\text{eq}(\hbar\omega+eV_\text{dc}+2n_1\hbar\omega_0) \nonumber \\
&+ J_{n_1-n}(\rho) S_\text{eq}(\hbar\omega-eV_\text{dc}-2n_1\hbar\omega_0)]
\end{align}
is non-symmetrized photon-assisted current noise \cite{mendes-mora-njp2015,grimsmo-blais-prl2106,mora-portier-prb2017}. It is defined in terms of the Bessel function $J_n(\rho = eV_\text{ac}/2\hbar\omega_0)$ and of the equilibrium current noise $S_\text{eq}(\omega) = 2e \langle \hat{I}_\text{qp}^i[\omega] \rangle /[1- \exp{(-\hbar \omega/k_\text{B} T)}]$ \cite{ingold-nazarov-book1992}.

Equation~\eqref{current} describes the linear response of the junction due to TL-voltage fluctuations. In the absence of ac-voltage, the linear response is strictly local in frequency with $Y_{n\neq 0}(\omega-2n\omega_0) = 0$. On the other hand, the addition of the ac bias leads to $Y_{n\neq 0}(\omega-2n\omega_0)\neq 0$ due to TL-voltage fluctuations at frequencies of the pump sidebands $2n \omega_0$. It is important to mention that $Y_{n\neq 0}(\omega-2n\omega_0)$ is only non-zero for nonlinear junctions \cite{mora-portier-prb2017}. Consequently, in this device, parametric interaction is due to the combination of both photon-assisted transport and nonlinearity. For instance, the nonlinearity converts a single pump photon of frequency $2\omega_0$ into two photons of frequency $\omega_0$ \cite{clerk-schoelkopf-rmp2010}. In the SIS amplifier, this process is characterized by $Y_1(\omega-2\omega_0)$ and allows for parametric amplification and squeezing \cite{mendes-mora-njp2015}. As will be clear from Eq.~\eqref{reflcoef}, parametric amplification arises when $\vert Y_1(-\omega_0)\vert\approx \vert Z_0^{-1} + Y_0(\omega_0) - iC\omega_0\vert$. Fig.~\ref{fig1b} illustrates the dc-voltage dependence of $Y_1(-\omega_0)$ and $Y_0(\omega_0)$ for a weak pump amplitude $eV_\text{ac}=0.155 \times 2\hbar\omega_0$. From the above criteria, parametric amplification and squeezing occurs for $eV_\text{dc} \approx 2\Delta-2\hbar\omega_0$ (see inset Fig.~\ref{fig1b}). At this dc voltage, single-photon dissipation Re$Y_0(\omega_0)$ is approximately zero, which is a necessary condition to reach near quantum-limited noise and large degree of squeezing by virtue of the fluctuation-dissipation theorem. Unsurprisingly, this dc voltage is also the optimal operational point of SIS mixers \cite{trucker-feldman-rmp1985}. In Fig.~\ref{fig1b}, the admittance present discontinuities for dc-voltages $eV_\text{dc} = 2\Delta-n\hbar\omega_0$. These discontinuities are a characteristic of photon-assisted transport and nonlinearity. The real parts exhibit jumps which are replica of the SIS $I$-$V$ curve while the imaginary parts exhibit logarithmic singularities. In practice, these discontinuities are rounded by experimental noise, finite Dynes parameter \cite{pekola-tsai-prl2010}, finite temperature and, in our simulations, by numerical precision. Finally, it is important to emphasize that $\hat{I}_\text{qp}[\omega]$ and $Y_{n}(\omega- 2n \omega_0)$ depend only on the quasiparticles dynamics determined by $H_\text{qp} + H_\text{qp}^\text{t}(t)$, and on the intrinsic characteristics of the junction, e.g., the superconducting gap $\Delta$ and normal state resistance $R_T$.

As mentioned above, parametric interaction emerges from two-photon processes characterized by $Y_{1}(\omega- 2\omega_0)$, while single-photon processes are characterized by $Y_{0}(\omega)$. The other non-local frequency contributions, from the second term of Eq.~\eqref{current}, give rise to conversion processes where photons of frequency $|\omega| > 2\omega_0$ are up- or down-converted to $\omega$. As theses contributions are detrimental to amplification and squeezing, we consider that an on-chip low-pass filter is used to filter all frequencies above $2\omega_0$. As a consequence, we can safely neglect all the contributions to current operator originating from $n \neq {0,1}$ in the sum over $n$ in Eq.~\eqref{current}, which takes the form
\begin{equation} \label{current0}
\hat{I}_\text{J}[\omega] = \hat{I}_\text{qp}^i[\omega] - Y_{0}(\omega) \hat{V}_\text{TL}^i[\omega ]- Y_{1}(\omega- 2\omega_0) \hat{V}_\text{TL}^i[\omega - 2\omega_0].
\end{equation}
In addition to the low-pass filter, the $R_T C$ time of the junction acts as a high-frequency cutoff. For the results presented in the next section, we take $\omega_{RC} = 1/R_T C = 2\pi \times 30$ GHz as a fixed parameter of the junction, which is obtained for a normal state resistance $R_T = 50~\Omega$ and $C=100$ fF.

Within these approximations, the expression for the outgoing field in terms of the incoming field and of the quasiparticles current operators is obtained by replacing Eq.~\eqref{current0} into Eq.~\eqref{inout0}. After simple algebraic manipulation, the outgoing field is
\begin{align} \label{inout1}
\an{out}[\omega_0 + \omega^\prime] & = r(\omega^\prime)\an{in}[\omega_0+ \omega^\prime] + \gamma(\omega^\prime)\ad{in}[\omega_0- \omega^\prime]\nonumber \\
&+ \alpha(\omega^\prime) \iqp [\omega_0+ \omega^\prime] + \beta(\omega^\prime)\iqp [ \omega^\prime-\omega_0]
\end{align}
with $\omega^\prime = \omega - \omega_0$ the frequency detuning and the reflection coefficient
\begin{equation} \label{reflcoef}
r(\omega^\prime) = \frac{[Z_0^{-1} - Y_J(\omega_s)][Z_0^{-1} + Y_J^*(\omega_i)] + \Gamma(\omega)}{[Z_0^{-1} + Y_J(\omega_s)][Z_0^{-1} + Y_J^*(\omega_i)] -  \Gamma(\omega^\prime)},
\end{equation}
where we defined $Y_J(\omega) = Y_0(\omega) - iC\omega$, $\omega_{s,i} = \omega_0\pm \omega^\prime$ and $\Gamma(\omega^\prime) = Y_1(\omega^\prime-\omega_0)Y_1^*(-\omega^\prime-\omega_0)$. Here, terms proportional to $Z_0^{-1}$ and $\text{Re}Y_{0}(\omega)$ are the internal TL decay rates at which the junction absorbs and emits photons, respectively. The terms proportional to $C$ and Im$Y_{0}(\omega)$ are related to the geometrical capacitance and dynamical susceptance of the junction. In the second term of Eq.~\eqref{inout1}, we have introduced $\gamma(\omega^\prime) = 2 \sqrt{\omega_i/\omega_s} Y_1(\omega^\prime-\omega_0)/Z_0\Delta_{\omega^\prime} $, with $\Delta_{\omega^\prime} = [Z_0^{-1} + Y_J(\omega_s)][Z_0^{-1} + Y_J^*(\omega_i)] -  \Gamma(\omega^\prime)$. As expected, the Fourier coefficient of $\ad{in}$ is proportional to $Y_1$. Moreover, the coefficients of $\iqp[\omega^\prime\pm\omega_0]$ are $\alpha(\omega^\prime)=  \mu_{\omega_s} [Z_0^{-1} + Y_J^*(\omega_i)]$, and $\beta(\omega^\prime) = \mu_{\omega_s}^* Y_1(\omega^\prime-\omega_0)$, with $\mu_{\omega_s}= i/\Delta_{\omega^\prime} \sqrt{\pi Z_0 \hbar \omega_s}$.

Equation~\eqref{inout1} is a central result of this paper and shows that the dc- and ac-voltage biased SIS junction acts as an amplifier operating in reflection mode. More specifically, for a fixed frequency detunig $\omega^\prime \neq 0$, the operators $\an{in}[\omega^\prime+\omega_0]$ and $\ad{in}[\omega_0-\omega^\prime]$ commute and can be seen as the input signal and the idler mode operators of a phase-preserving amplifier \cite{caves-prd1982,clerk-schoelkopf-rmp2010}. In this mode, the gain is simply given by $G(\omega^\prime) = |r(\omega^\prime)|^2$ and the last two terms of Eq.~\eqref{inout1} correspond to added noise beyond the quantum limit $\mathcal{A}_\text{q}(\omega^\prime) =[1-1/G(\omega^\prime)]/2$. To characterize the effects of quasiparticle tunneling on the performance of the amplifier, we compute the added noise using the relation $\langle |\an{out}[\omega]|\rangle = G(\omega)(\mathcal{A}(\omega)  + \langle |\an{in}[\omega]|\rangle)$ \cite{caves-prd1982,boutin-blais-prappl2017}, with $\langle |\hat{O}|\rangle = \langle \hat{O}^\dagger \hat{O} + \hat{O}\hat{O}^\dagger \rangle/2$ and the added noise
\begin{align} \label{addednoise1}
\mathcal{A}(\omega^\prime) &=  \mathcal{A}_\text{q}(\omega^\prime) + \frac{2\pi}{G(\omega^\prime)} \left[ |\alpha(\omega^\prime)|^2 S_0(\omega_s) \nonumber \right.  \\
&\left.  + |\beta(\omega^\prime)|^2 S_0(-\omega_i) - 2\re [\alpha(\omega^\prime)\beta^*(\omega^\prime)] S_1(\omega_s) \right].
\end{align}
The terms proportional to $S_0(\pm \omega_{s,i})$ and $S_1(\omega_s,)$ are noise generated by the tunneling of quasiparticles, and they originate from absorption and emission processes of one and two quantum of energy $\hbar\omega_0$ by the junction. As expected, the quasiparticle noise $S_n(\omega)$ increases the added noise of the amplifier. Thus, to mitigate the effects of the quasiparticle noise, the operational voltages of the device are such that $S_0(\pm\omega_{s,i})$ and $S_1(\omega_s)$ are as small as possible. To characterize the deviation of the quantum limit, we define the quantum efficiency of parametric amplifier as $\xi = \mathcal{A}_\text{q}(\omega)/\mathcal{A}(\omega)$. For an ideal phase-preserving amplifier $\xi = 1$ and in the presence of quasiparticle noise $\xi < 1$.

For $\omega^\prime = 0$, $\an{in}[\omega_0]$ and $\ad{in}[\omega_0]$ do not commute, and the first two terms of Eq.~\eqref{inout1} rather characterize an ideal phase-sensitive amplifier \cite{caves-prd1982,clerk-schoelkopf-rmp2010}. In this operational mode, the gain $G(0)$ is defined as a combination of $r(0)$, $\gamma (0)$ and the phases of the input and output fields \cite{boutin-blais-prappl2017}. The added noise is given by a linear combination of the two last terms of Eq.~\eqref{inout1} and it also depends on the input and output field phases \cite{boutin-blais-prappl2017}.

To better characterize the radiation emitted by the SIS amplifier, we also compute bellow the output power spectrum $S_\theta(\omega)$, defined in terms of $\langle \Delta \hat{X}_{\theta}[\omega]  \Delta \hat{X}_{\theta}[\omega_1] \rangle = S_\theta(\omega) \delta(\omega+\omega_1)$, with $\hat{X}_{\theta}[\omega^\prime] = e^{-i \theta}\hat{a}_\text{out}[\omega^\prime+\omega_0] + e^{i \theta}\hat{a}_\text{out}^{\da}[\omega_0 -\omega^\prime]$ the output field quadrature with fluctuations $\Delta \hat{X}_{\theta}[\omega] = \hat{X}_{\theta}[\omega] - \langle \hat{X}_{\theta}[\omega]  \rangle$ and $\theta$ the phase of the output field. Similarly to amplification, the emitted radiation can be also characterized by single-mode $S_\theta(\omega^\prime=0) < 1$ or two-mode $S_\theta(\omega^\prime \neq 0) < 1$ squeezed state.

\section{Results \label{results}}

We first present results for an ideal SIS junction, with transport response rising steeply for voltages $eV_\text{dc} = 2\Delta - n\hbar \omega_0$ as illustrated in Fig.~\ref{fig1b}(b). We then investigate how gain and squeezing properties are affected by low-frequency noise, which are smoothing out the transport response of the junction and, consequently, diminishing the strength of the parametric interaction. Lastly, an impedance matching scheme is presented. This scheme is used to match both the geometrical capacitance $(C)$ and dynamical susceptance [$\im Y_0(\omega)$] of the junction, thus leading to much larger bandwidths.

\subsection{Ideal SIS junction \label{resultsA}}

\begin{figure}[t]
\begin{center}
\includegraphics[width=0.45\textwidth]{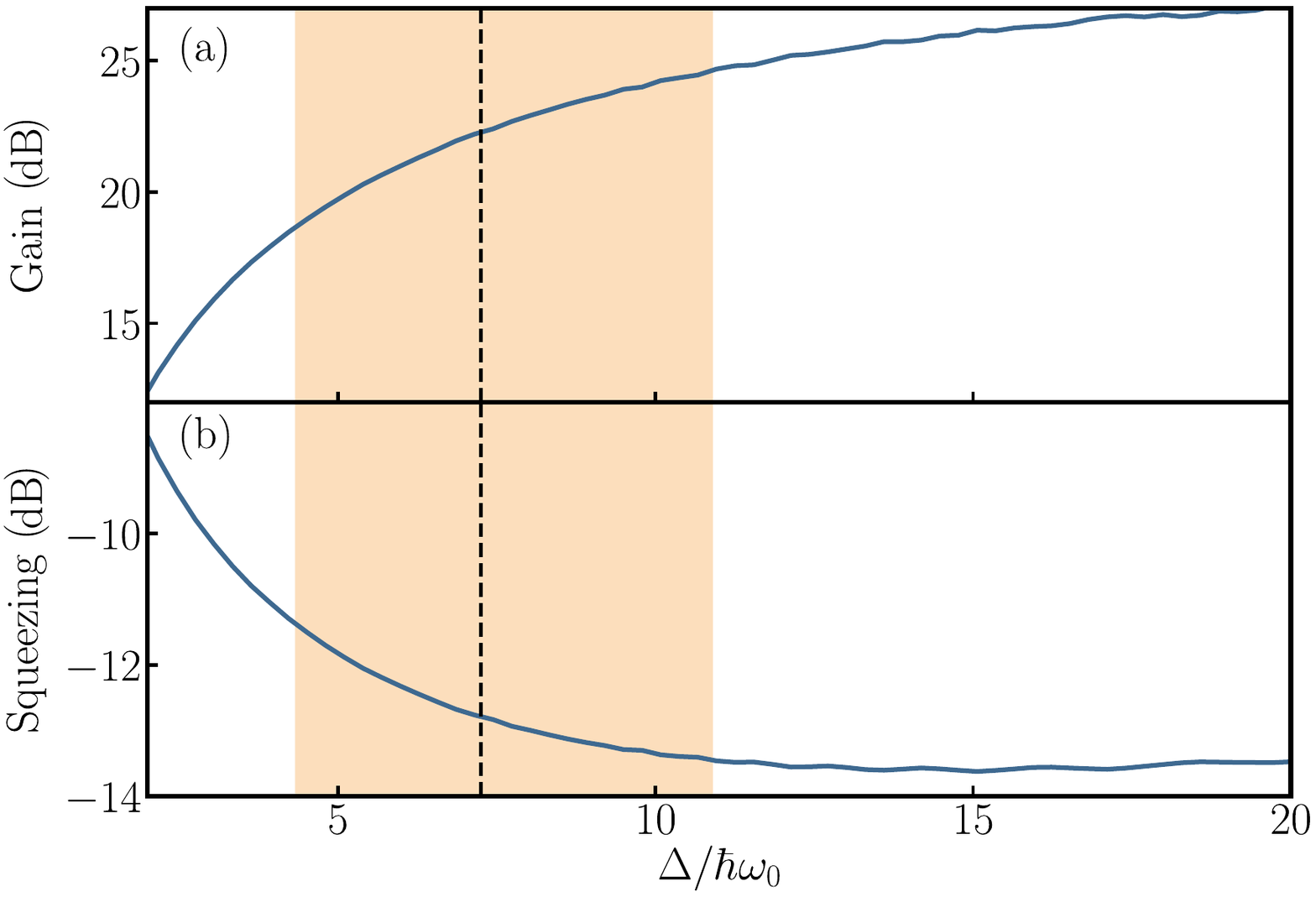}
\caption{(a) Phase-sensitive gain and (b) single-mode squeezing as a function of $\dw$ for $\omega_0/2\pi = 6$ GHz. The filled area corresponds to values for an aluminum superconducting junction ($\Delta = 180 \mu eV$) for $\omega_0/2\pi$ in the 4 to 10 GHz, and the dashed line marks $\omega_0/2\pi = 6$ GHz used to investigate the frequency-dependent gain and squeezing. For each value of $\dw$, the dc voltage is fixed to $eV_\text{dc} = 2\Delta - 1.999\hbar \omega_0$ and the ac voltage is optimized to maximize squeezing. The optimal ac-voltage amplitude diminishes as $\dw$ increases (not shown here). \label{fig2}} 
\end{center}
\end{figure}

\begin{figure}[t]
\begin{center}
\includegraphics[width=0.45\textwidth]{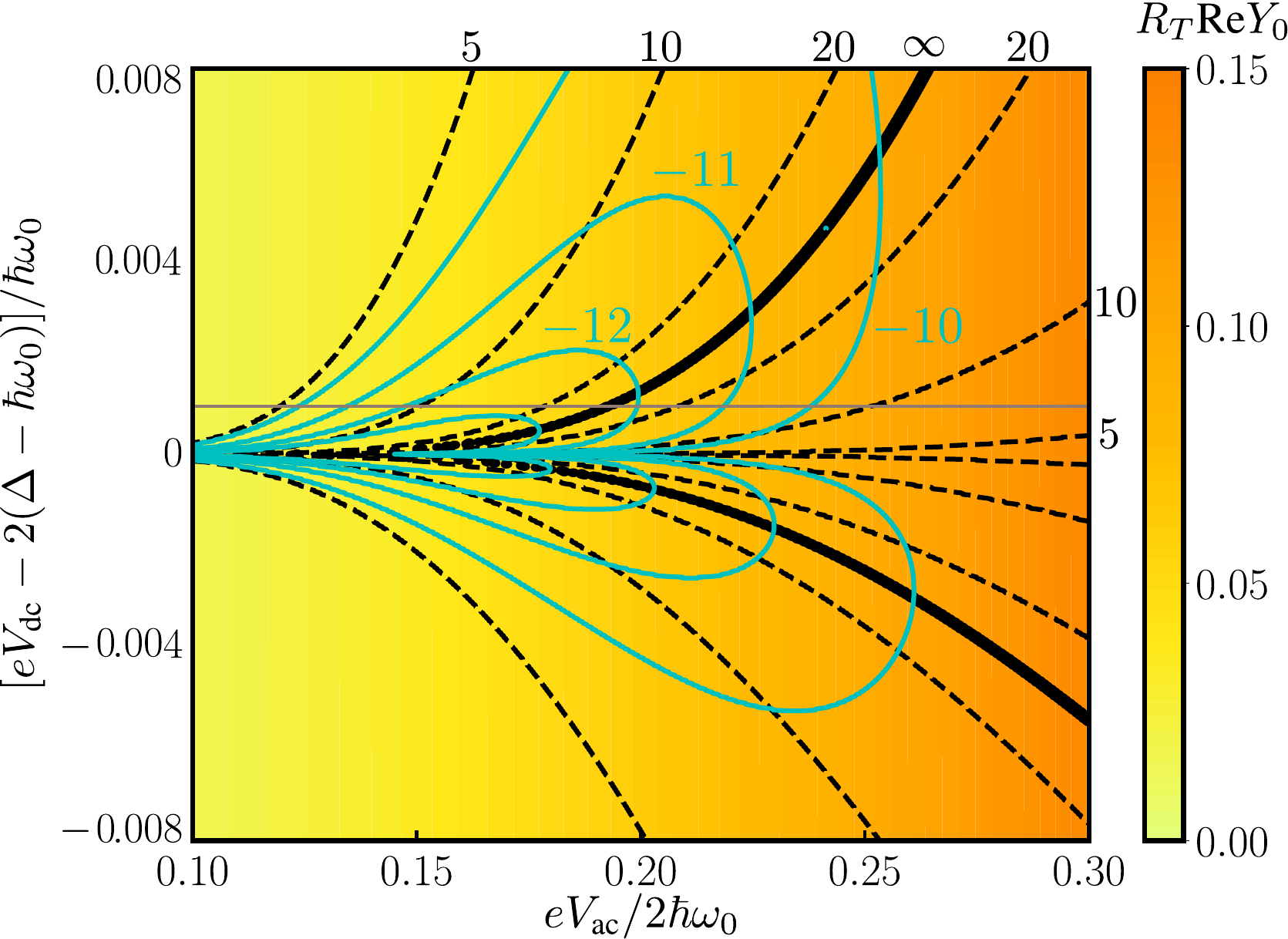}
\caption{DC- and AC-voltage response of an aluminum SIS junction with $\omega_0/2\pi = 6$ GHz ($\dw\approx 7.2$) and $\omega^\prime=0$: single-photon dissipation $\re Y_0$ (heatmap), phase-preserving gain (dashed contours) and squeezing (solid contours) near the logarithmic singularity $eV_\text{dc}=2(\Delta-\hbar\omega_0)$. Gains from 5 to 20~dB and squeezing betwen $-10$ and $-13$ are represented by, respectively, constant values indicated in dB next to the contours. The thick solid line indicates voltages where the phase-preserving gain becomes extremely large. \label{fig3}}
\end{center}
\end{figure}

Before considering the general frequency-dependent amplification and squeezing, we present results for zero frequency detuning ($\omega^\prime = 0$). First, we investigate the effect of the strength of nonlinearities on the phase-sensitive mode. Figs.~\ref{fig2}(a) and (b) illustrate, respectively, the phase-sensitive gain and single-mode squeezing as a function of $\dw$, when optimizing $V_\text{ac}$ to maximize single-mode squeezing for $eV_\text{dc} = 2\Delta - 1.999\hbar\omega_0$ (see horizontal solid line in Fig.~\ref{fig3}). This dc voltage is in the vicinity of the logarithmic singularity showed in the inset of Fig.~\ref{fig1b}; its choice is explained Fig.~\ref{fig3}. In Fig.~\ref{fig2}, $\Delta$ varies while $\omega_0/2\pi$ is kept fixed and equal to 6 GHz. Already at small $\dw \simeq 2$ we observe more than $\sim 10$ dB of amplification and $\sim -8$ dB of squeezing. As $\dw$ increases, the strength of the nonlinearities giving rise to parametric interaction increases, thus enhancing gain and squeezing as illustrated in Fig.~\ref{fig2}. Thus, increasing $\Delta$ leads to an increase of the parametric interaction strength $\propto Y_1(\omega)$. For $\dw = 20$, amplification and squeezing reach approximately $27$ dB and $-13.5$ dB, respectively. Here, the filled area corresponds to $\omega_0/2\pi$ in the 4 to 10 GHz range for an aluminum junction with $\Delta/h \approx 43.5$ GHz. A larger value of $\Delta$ can be obtained by reducing the thickness of the aluminum layer \cite{chubov-pilipenko-jetp1969} or with a different superconductor. The dashed line corresponds to $\omega_0/2\pi =6$ GHz, the value used to investigate the frequency-dependent features of the SIS amplifier. The optimal ac-voltage decreases as $\dw$ increases (not shown).

For the remainder of this article, we consider an aluminum SIS junction with $\omega_0/2\pi = 6$ GHz ($\dw \approx 7.2$). To further gain insight on the parametric amplifier operational voltages, we investigate dc- and ac-voltage dependence of single-mode squeezing and phase-preserving gain at zero frequency detuning $\omega^\prime=0$. Fig.~\ref{fig3} shows contours of constant gain (dashed lines) and squeezing (solid lines) in the vicinity of the logarithmic singularity $eV_\text{dc}=2(\Delta-\hbar\omega_0)$.

We first investigate the phase-preserving gain. The different voltage points realizing a given gain do not result in the same device performances, which are characterized by quantum efficiency and 3 dB bandwidth. The former improves when the quasiparticle noise terms in Eq.~\eqref{addednoise1} are reduced. By virtue of the fluctuation-dissipation theorem, noise is related to the dissipation $\re Y_0$. Fig.~\ref{fig3} shows that, in the voltage range of interest, $\re Y_0$ is nearly independent of dc-voltage and increases with increasing ac voltage. Accordingly, along a contour of constant gain, going to lower ac voltage increases the quantum efficiency. Remarkably, we find numerically that it also increases the 3 dB bandwidth. In principle, in the limit $V_{ac}\rightarrow0^+$, one would get perfect quantum efficiency, $\xi=1$. However, for a given gain, decreasing the ac voltage requires to set the dc voltage increasingly closer to the logarithmic singularity. At some point, the logarithmic singularity is smoothed out by experimental noise on the dc voltage or non-ideal $I$-$V$ curve (finite Dynes parameter). This will set the best experimental performances of the device. For the remainder of the paper, the dc voltage is set to the experimentally relevant value $eV_\text{dc} = 2\Delta-1.999\hbar\omega_0$ and, consequently, gain and squeezing are tuned in-situ by varying the ac-voltage $V_\text{ac}$ (horizontal line in Fig.~\ref{fig3}).

We now turn to squeezing. Similarly to quantum efficiency, squeezing is strongly sensitive to the quasiparticle noise terms, which are related to $\re Y_0$. For a given $V_\text{ac}$, $\re Y_0$ is almost constant. As a consequence, we expect squeezing to be maximum for the $V_\text{dc}$ giving infinite gain (thick solid curve in Fig.~\ref{fig3}, where the denominator of the gain, $G(\omega) = |r(\omega)|^2$, tends to zero, i.e, $\vert Y_1(-\omega_0)\vert = \vert Z_0^{-1} + Y_0(\omega_0) - iC\omega_0\vert$.). At infinite gain, an ideal amplifier would also produce infinite squeezing; however, the finite quasiparticle noise (Re$Y_0$) bounds squeezing. Analogously to quantum efficiency, maximum squeezing will thus continuously improve by lowering the ac voltage at the expense of setting the dc voltage closer to the logarithmic discontinuity. Also infinite squeezing could be achieved in the limit $V_\text{ac}\rightarrow0^+$, the maximum experimentally achievable value will be bounded by the smoothing of the logarithmic singularity. Squeezing reaching -15~dB could be achieved with the current experimental setups.

\begin{figure}[t]
\begin{center}
\includegraphics[width=0.45\textwidth]{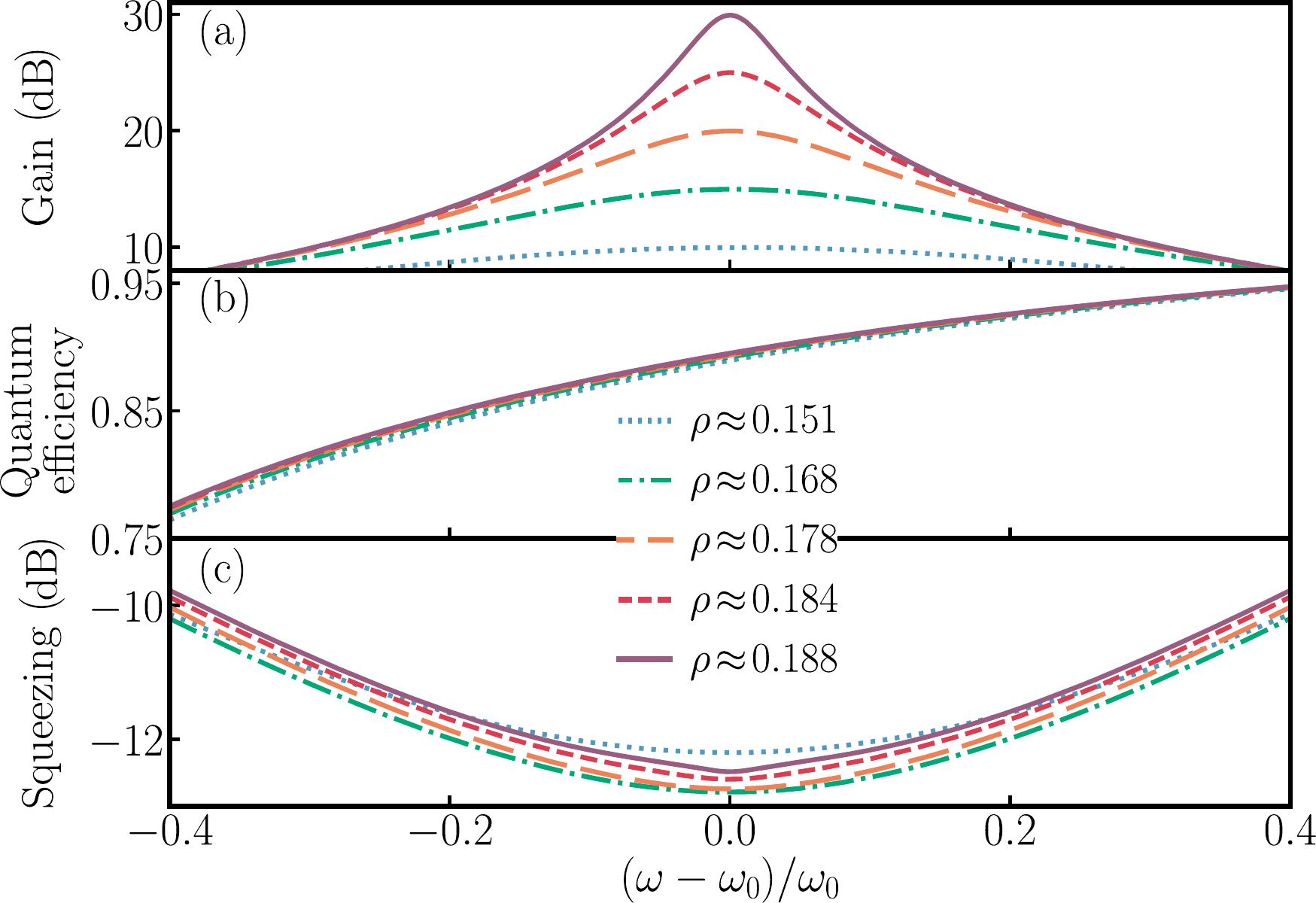}
\caption{Aluminum SIS junction with $\omega_0/2\pi = 6$ GHz ($\dw\approx 7.2$): Phase-preserving gain (a), quantum efficiency (b) and squeezing (c) as a function of the frequency detuning from $\omega_0$ for the dc voltage $eV_\text{dc} = 2\Delta-1.999\hbar\omega_0$ and five different values of ac-voltage amplitude, $\rho = eV_\text{ac}/2\hbar \omega_0 \approx 0.151$ (dotted line), $0.168$ (dashed-dotted line), $0.178$ (long-dashed line), $0.184$ (dashed line), $0.188$ (solid line). These values of the ac voltage are chosen such that the phase-preserving gain is, respectively, equal to 10, 15, 20, 25 and 30~dB at zero frequency detuning $\omega=0$. \label{fig4}}
\end{center}
\end{figure}

Figure~\ref{fig4} shows (a) phase-preserving gain, (b) quantum efficiency and (c) squeezing as a function of frequency detuning for $eV_\text{dc}=2\Delta-1.999\hbar\omega_0$ and several ac-voltage amplitudes: $\rho = eV_\text{ac}/2\hbar \omega_0 \approx 0.151$ (dotted line), $0.168$ (dashed-dotted line), $0.178$ (long-dashed line), $0.184$ (dashed line), $0.188$ (solid line). The gain, Fig.~\ref{fig4}(a), is observed to increase with increasing ac-voltage amplitude while its 3 dB bandwidth decreases. In the 20 to 30 dB range, the gain-bandwidth product is approximately constant, equal to 12 GHz. This value is 87\% larger than the gain-bandwidth product of the broadband impedance engineered JPA \cite{roy-vijay-apl2015}. Fig.~\ref{fig4}(b) shows that the quantum efficiency is close to 0.9 in the 3 dB bandwidth frequency range and nearly independent of ac-voltage amplitude. This is expected from the small but finite $\re Y_0$ observed in Fig.~\ref{fig3}. Fig.~\ref{fig4}(c) illustrates a far-separated two-mode squeezing with 3 dB bandwidth reaching $\sim 5.3$~GHz. Moreover, unlike gain, squeezing varies only weakly with the ac-voltage amplitude and, after it reaches its maximum value, further increase of the ac-voltage amplitude reduces squeezing, as expected from Fig.~\ref{fig3}.

\subsection{Effects of low-frequency noise \label{resultsB}}

The results presented in the previous section were obtained considering an ideal SIS junction in the low-temperature limit $k_\text{B} T \ll 2\Delta$ for which the transport response is singular and discontinuous for $eV_\text{dc}=2\Delta-n\hbar\omega_0$ \cite{barone-paterno-book}. However, this is an idealized situation and, in practice, temperature or low-frequency noise can smoothen the transport response. Here, we consider the effects of low-frequency noise on gain and squeezing properties of the SIS amplifier. These effects are included by assuming that the junction interacts with a low-frequency electromagnetic environment \cite{hofheinz-esteve-prl2011} and that the transport properties are described by the $P(E)$-theory \cite{ingold-nazarov-book1992}. This approach has been shown to quantitatively explain the finite Dynes tunneling density of states \cite{pekola-tsai-prl2010}, usually observed below the dc-transport gap in a normal-insulator-superconductor junctions and the corresponding smoothing of the BCS coherence peak \cite{barone-paterno-book}. In this approach, the low-frequency noise modifies the equilibrium noise current noise to 
\begin{equation} \label{noiseeq-pe}
S_\text{eq}^\text{eff}(\omega) = \int_{-\infty}^{\infty} S_\text{eq}(\hbar\omega - E) P(E)  dE,
\end{equation}
where $P(E)$ is the probability density of a tunneling quasiparticle emitting energy $E$ \cite{ingold-nazarov-book1992}.

To model the low-frequency electromagnetic environment, we consider that it originates from the dc-bias scheme. In general, the biasing scheme consists of a resistive voltage divider followed by large capacitive filtering \cite{hofheinz-esteve-prl2011,chen-rimberg-prb2014}. In this situation, the low-frequency impedance is thus the parallel combination of a resistance with a large capacitance $C_{bt}$. This filtering scheme reduces the bandwidth over which low-frequency voltage noise is detrimental to the kHz range \cite{chen-rimberg-prb2014}, making the low-frequency voltage fluctuations fully classical. In this setup, the effect low-frequency environment on the transport properties is then described by
\begin{equation}
P(E) = \frac{1}{\sqrt{4\pi E_c  k_\text{B} T}}\exp \left( - \frac{(E-E_c)^2}{4  E_c   k_\text{B} T} \right) ,
\end{equation}
with $E_c = e^2/2C_{bt}$ is the capacitor charging energy. With this model, the theory developed in Sec.~\ref{model} remains the same except for the replacement of $S_\text{eq}(\omega)$ by $S_\text{eq}^\text{eff}(\omega)$ in Eq.~\eqref{noise-PE}.
\begin{figure}[t]
\begin{center}
\includegraphics[width=0.45\textwidth]{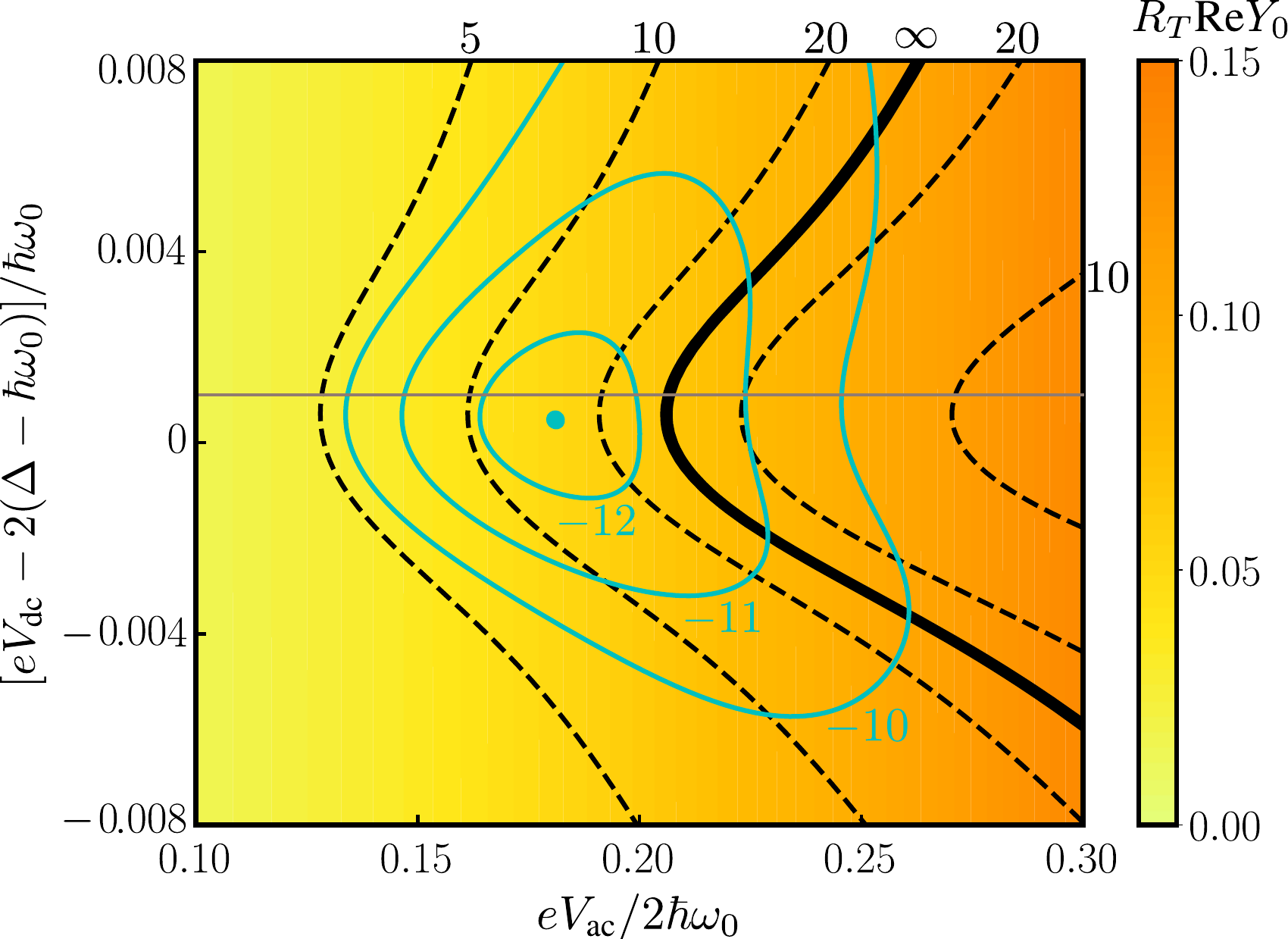}
\caption{DC- and AC-voltage response of an aluminum SIS junction with $\omega_0/2\pi = 6$ GHz ($\dw\approx 7.2$) and $\omega^\prime=0$: single-photon dissipation $\re Y_0$ (heatmap), phase-preserving gain (dashed contours) and squeezing (solid contours) near the logarithmic singularity $eV_\text{dc}=2(\Delta-\hbar\omega_0)$ in the presence of low-frequency environment. The low-frequency noise filtering scheme is characterized by the capacitor $C_{bt} = 100$ pF which results in rms voltage fluctuations of 35 nV for $T=15$ mK. Gains from 5 to 20~dB and squeezing between $-10$ and $-12$ are represented by, respectively, constant values indicated in dB next to the contours. The round-blue dot indicates the maximum squeezing, $\sim -12.3$ dB, voltage point. The thick solid line indicates voltages where the phase-preserving gain becomes extremely large. The horizontal solid line marks the dc-voltage $eV_\text{dc} = 2\Delta - 1.999\hbar\omega_0$.
\label{fig5.0}}
\end{center}
\end{figure}

Figure~\ref{fig5.0} illustrates the effect of low-frequency environment on gain (dashed lines), squeezing (solid lines) and the single-photon dissipation Re$Y_0$ (heat map) as a function of the dc- and ac-voltages. The filtering scheme is characterized by $C_{bt} = 0.1$~nF corresponding to rms voltage fluctuations of $\sqrt{k_\text{B} T/2C_{bt}} \sim 35$~nV. As expected, the low-frequency noise removes the logarithmic singularity at $eV_\text{dc} = 2(\Delta -\hbar \omega_0)$ (see Fig.~\ref{fig3}). Similarly to the ideal case, single-photon dissipation Re$Y_0$ is nearly independent of the dc voltage and increases with increasing ac voltage. Maximization of the device performances will thus again require to lower the ac voltage. However, along a contour of constant gain (dashed lines) or squeezing (solid lines), $V_\text{ac}$ now has a finite lower bound. Unlike Fig.~\ref{fig3} where the device performances constantly increased toward $\xi=1$ and infinite squeezing by setting the dc voltage closer and closer to the logarithmic singularity $eV_\text{dc}=2(\Delta-\hbar\omega_0)$, here we expect approximately constant performances when $2(\Delta-\hbar\omega_0)\lesssim eV_\text{dc}\lesssim 2\Delta-1.999\hbar\omega_0$. The presence of low-frequency noise bounds the squeezing to a maximum value of approximately $-12.3$~dB (round-blue dot). The maximum squeezing depends strongly on the filtering capacitor $C_{bt}$ and it decreases for smaller capacitances. In contrast, note that gain remains tuneable to any value up to infinity by the dc- and ac-voltages at the expense of increasing single-photon dissipation Re$Y_0$.
\begin{figure}[t]
\begin{center}
\includegraphics[width=0.45\textwidth]{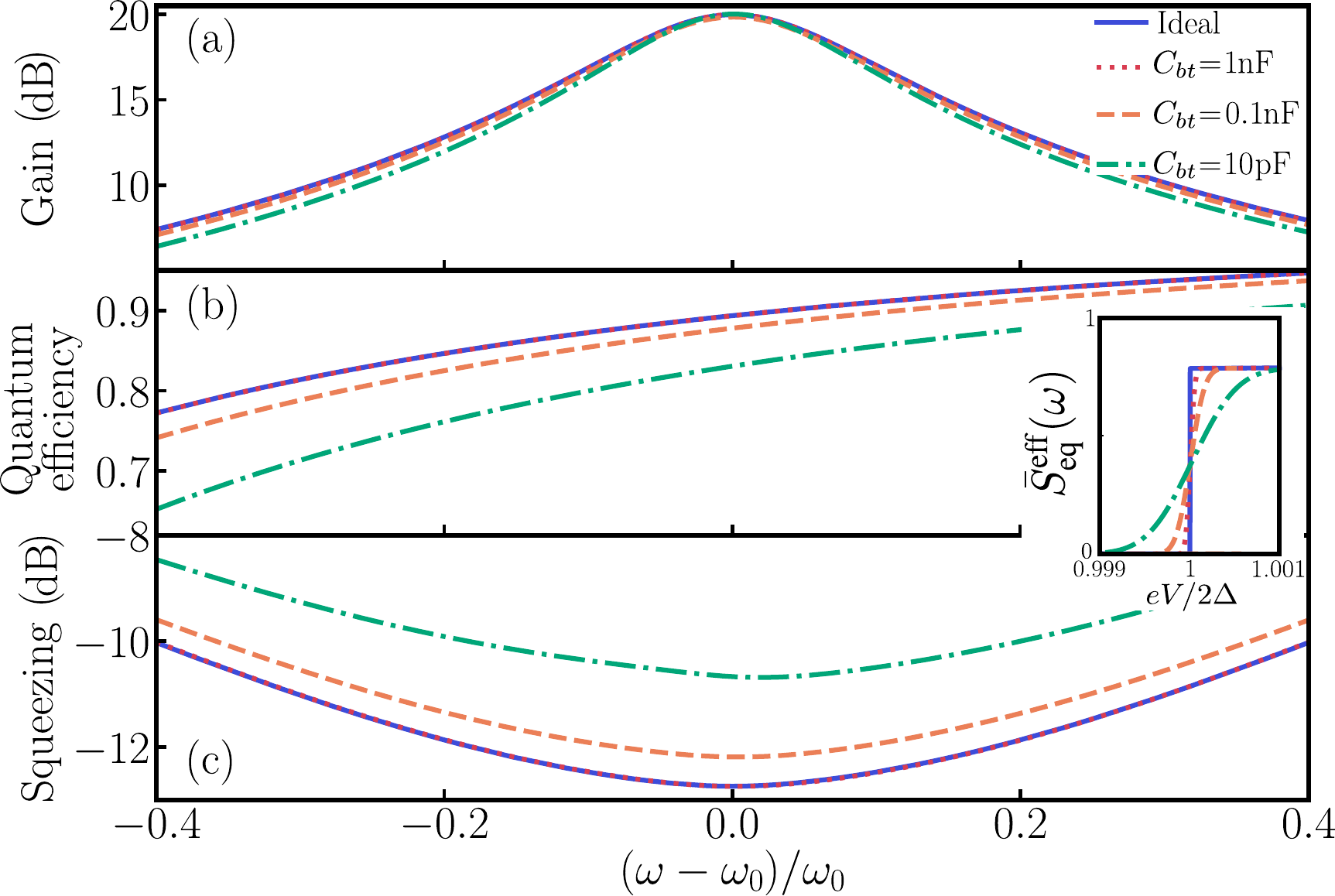}
\caption{Aluminum junction ($\Delta/\hbar\omega_0 \approx 10.9$): Gain (a), added photon number (b) and squeezing (c) as a function of frequency detuning in the presence and in the absence (solid line) of low-frequency electromagnetic environment. The dotted line illustrates an efficient ($C_{bt}\sim 1$~nF) filtering scheme. On the other hand, the dashed ($C_{bt}\sim 0.1$~nF) and the dashed-dotted ($C_{bt}\sim 10$~pF) lines represent a less efficient low-frequency noise filtering. The voltages were chosen such that the gain at zero detuning is equal to 20 dB for all filtering capacitances. As expected, an efficient filtering scheme leads to gain, added noise and squeezing similar to the ideal case. Moreover, the gain bandwidth is only weakly affected by the low-frequency electromagnetic environment. However, the amplitude of quantum efficiency and squeezing are dimineshed by low-frequency noise. The inset illustrates the effect of the low-frequency environment on equilibrium noise $\bar{S}_\text{eq}^\text{eff}(\omega)$ in units of $2\Delta/R_T$. \label{fig5.5}}
\end{center}
\end{figure}

We now investigate the effects of the low-frequency noise on the frequency-dependent properties of the amplifier. Fig.~\ref{fig5.5} illustrates (a) gain, (b) quantum efficiency and (c) squeezing for $eV_\text{dc} = 2\Delta - 1.999\hbar\omega_0$ and three different filtering capacitances: $C_{bt}\sim 1$~nF (dotted line), $C_{bt}\sim 0.1$~nF (dashed line) and $C_{bt}\sim 10$~pF (dashed-dotted line), respectively. The ac voltage is set to give 20 dB of gain at zero detuning. These results are compared with the ideal case (solid line), where the junction does not interact with the low-frequency environment. For an aluminum junction ($\Delta/\hbar\omega_0 \approx 10.9$) and $T=15$~mK, $C_{bt}\sim 1$~nF corresponds to rms voltage fluctuations of $\sqrt{k_\text{B}T/2C_{bt}} \sim 11$~nV (dot-dashed line), the gain, added noise and squeezing are only weakly affected by the low-frequency environment which is efficiently filtered. The effect of low-frequency noise on $S_\text{eq}^\text{eff}(\omega)$ is illustrated in the inset of Fig.~\ref{fig5.5}. For $C_{bt} \sim 1$~nF, $S_\text{eq}^\text{eff}(\omega)$ is almost indistinguishable from the ideal case (dashed line), a signature that the low-frequency noise is efficiently filtered. On the other hand, under less efficient filtering, $C_{bt} \sim 10$~pF corresponding to $\sim 0.1~\mu$V rms voltage fluctuations (dashed line) and $C_{bt} \sim 0.1$~nF, the equilibrium current noise rises smoothly and its behavior near $2\Delta/\hbar$ deviates from the ideal case (see inset). In these cases, low-frequency noise diminishes both quantum efficiency and squeezing [dashed and dashed-dotted lines in (b) and (c)]. However, the 3 dB bandwidth is weakly diminished. Furthermore, the main effect of the low-frequency noise is to diminish the strength of the nonlinearity giving rise to parametric interaction and, therefore, to obtain higher gains the ac-voltage amplitude must be increased. The increasing of the ac voltage enhances single-photon dissipation Re$Y_0$ and degrades quantum efficiency and squeezing.

\subsection{Impedance matching \label{impeng}}

\begin{figure}[t]
\begin{center}
\includegraphics[width=0.45\textwidth]{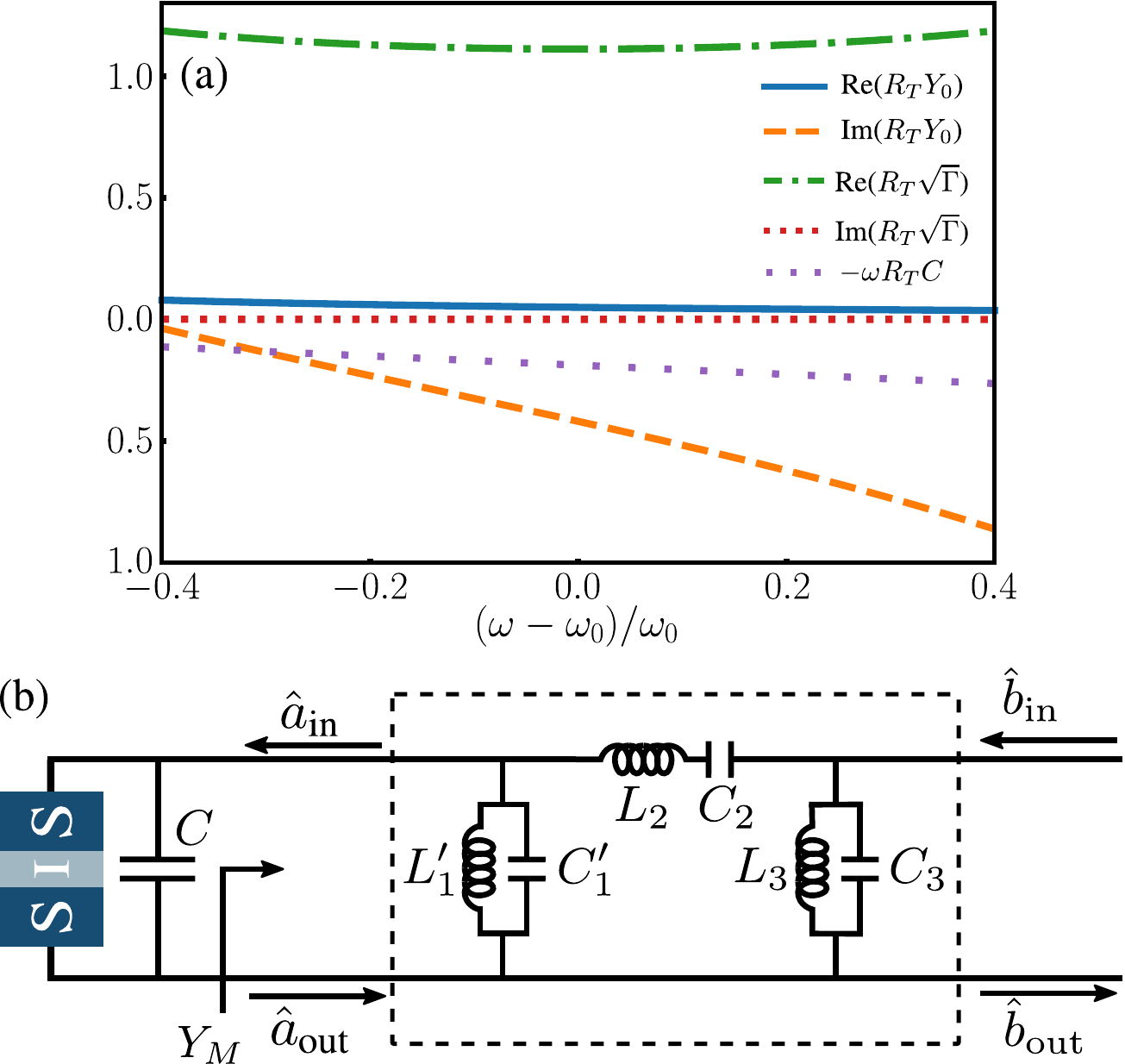}
\caption{(a) Frequency dependence of the terms appearing in the reflexion coefficient [Eq.~\eqref{reflcoef}]. Whereas the losses $\re Y_0$, the parametric two-photon process $\Gamma$ and the capacitance $\omega C$ are fairly flat, the dynamical susceptance $\im Y_0$ has strong frequency dependence. The later is modeled with reasonable accuracy as a parallel $LC$ tank, i.e, the bandpass filter structure of the matching network. The parameters used are: $\omega_0/2\pi = 6$ GHz, $\dw \approx 7.2$, $eV_\text{dc} = 2\Delta-1.999\hbar\omega_0$, $eV_\text{ac}/2\hbar \omega_0 \approx 0.178$ and $C=100$~fF; they correspond to a phase-preserving gain of $20$~dB at frequency $\omega=\omega_0$ in the absence of the matching network. (b) Electrical circuit matching the dynamical susceptance Im$Y_0$ and geometric capacitance $C$ of the junction. The input-output operators $\hat{a}_\text{in}$ and $\hat{a}_\text{out}$ obey the same equations as those in Fig.~\ref{fig1a}, in particular Eq.~(\ref{inout1}). However, the experimentally controllable and measurable observables are the input-output operators $\hat{b}_\text{in}$ and $\hat{b}_\text{out}$. The fields $\hat{a}$'s and $\hat{b}$'s are related by the $S$-matrix of the matching network (black box), see Eqs.~\eqref{Smatrix} and \eqref{Smatrixconj}. The matching network assumes a bandpass filter structure in order to compensate the junction dynamical susceptance $\im Y_0$ and geometric capacitance $C$. \label{fig6}}
\end{center}
\end{figure}

At first sight, the aforementioned gain-bandwidth product of 12~GHz is quite surprising since the energy scale of the SIS parametric amplifier is the superconducting gap voltage ($2\Delta$), which corresponds to frequencies of approximately 90 GHz for aluminum junctions. To understand what is limiting the gain-bandwidth product, we recall that the phase-preserving gain is determined by the reflection coefficient [Eq.~\eqref{reflcoef}], and its frequency dependence is due to $Y_J(\omega)$ and $\Gamma(\omega)$. Fig.~\ref{fig6}(a) illustrates the frequency dependence of $Y_J(\omega)$ and $\Gamma(\omega)$ for $eV_\text{dc} = 2\Delta-1.999\hbar\omega_0$ and $eV_\text{ac}/2\hbar\omega_0 = 0.178$. The parametric conversion term $\Gamma(\omega)$ is real, independently of the voltages, and is fairly flat in a wide band of frequencies. Also, as expected, the dissipation $\re Y_0$ is negligible for all frequencies. The contribution of the geometrical capacitance depends weakly on frequency. Consequently, the strong frequency dependence of gain originates from the dynamical susceptance $\im Y_0$.

In order to develop a matching scheme to increase the gain-bandwidth product, we first note that the frequency dependence of Im$Y_0(\omega)$ is modeled with reasonable accuracy by a parallel $L_\text{eff}C_\text{eff}$ tank with the following characteristics: $\omega_0R_TC_\text{eff}=0.7$ and $R_T/\omega_0L_\text{eff}=0.25$. Note that the precise values of $L_\text{eff}$ and $C_\text{eff}$ depend on $V_\text{dc}$ and $V_\text{ac}$. However, the above approximate values of $L_\text{eff}$ and $C_\text{eff}$ are sufficient to match Im$Y_0$ for a wide range of voltages. Following Ref.~\onlinecite{matthaei-book}, the matching scheme consists in adding, between the SIS junction and the TL, a bandpass filter network (black box). In the bandpass filter network, depicted in Fig.~\ref{fig6}(b), the first capacitance (inductance) $C_1^\prime$  ($L_1^\prime$) has been reduced (increased) compared to the original filter capacitance $C_1$ (inductance $L_1$) to $C_1^\prime = C_1-C_\text{eff}-C$ ($1/L_1^\prime = 1/L_1 - 1/L_\text{eff}$). In this manner, the effect of $\im Y_0$ and $C$ is  absorbed in the filter. Since the filter is designed to achieve good matching in the passband, the detrimental effects of $\im Y_0$ and $C$ are eliminated.

To obtain an experimentally useful parametric amplifier, we choose filter elements $L_n$ and $C_n$ giving a 3-poles Chebyshev filter with 4-8~GHz band edges. Such filters are also characterized by their ripples. While the Chebyshev filter benefically leads to an increase in the gain-bandwidht product, its ripples on the other hand impact directly the gain flatness. The ripples amplitude is directly related to the capacitance $C_1$ of the filter network. To reduce the amplitude of the ripples, the capacitance $C_1$ must be made as small as possible~\cite{matthaei-book} with the constraint $C_1^\prime >0$. In this way, we choose $C_1 \gtrsim C_\text{eff}+C$ to minimize the amplitude of the ripples.

Once the elements of the filter are determined, we can readily incorporate it in the quantum formalism via the $S$-matrix formalism, which gives the sets of linear equations
\begin{equation} \label{Smatrix}
\begin{pmatrix}
\hat{b}_\text{out}[\omega_0+\omega^\prime]\\
\hat{a}_\text{in}[\omega_0+\omega^\prime]
\end{pmatrix}
=
S(\omega_0+\omega^\prime)
\begin{pmatrix}
\hat{b}_\text{in}[\omega_0+\omega^\prime]\\
\hat{a}_\text{out}[\omega_0+\omega^\prime]
\end{pmatrix}
\end{equation}
and
\begin{equation} \label{Smatrixconj}
\begin{pmatrix}
\hat{b}_\text{out}^\dagger[\omega_0-\omega^\prime]\\
\hat{a}_\text{in}^\dagger[\omega_0-\omega^\prime]
\end{pmatrix}
=
S^*(\omega_0-\omega^\prime)
\begin{pmatrix}
\hat{b}_\text{in}^\dagger[\omega_0-\omega^\prime]\\
\hat{a}_\text{out}^\dagger[\omega_0-\omega^\prime]
\end{pmatrix},
\end{equation}
to be solved together with Eq.~\eqref{inout1}. Here, $S(\omega)$ is the $S$-matrix that connects the SIS junction fields $\hat{a}_\text{in/out}$ to the TL fields $\hat{b}_\text{in/out}$ [see Fig.~\ref{fig6}(b)]. With this formalism, the definitions of the gain, quantum efficiency and squeezing remain the same provided that $\hat{a}_\text{out}$ is replaced by $\hat{b}_\text{out}$. In the presence of the filter, the phase-preserving gain takes the form~\cite{matthaei-paramp}
\begin{equation} \label{gain-Y-matching}
G(\omega^\prime) = \left\vert 
	\frac{\left[Y^*_M(\omega_s) - Y_J(\omega_s)\right] \left[Y^*_M(\omega_i) + Y^*_J(\omega_i)\right] + \Gamma(\omega^\prime)}
	{\left[Y_M(\omega_s) + Y_J(\omega_s)\right] \left[Y^*_M(\omega_i) + Y^*_J(\omega_i)\right] - \Gamma(\omega^\prime)}
	\right\vert^2,
\end{equation}
where $\omega_{s,i} = \omega_0 \pm \omega^\prime$ and $Y_M$ is the filter-matching-network admittance [see Fig.~\ref{fig6}(a)].

\begin{figure}[t]
\begin{center}
\includegraphics[width=0.45\textwidth]{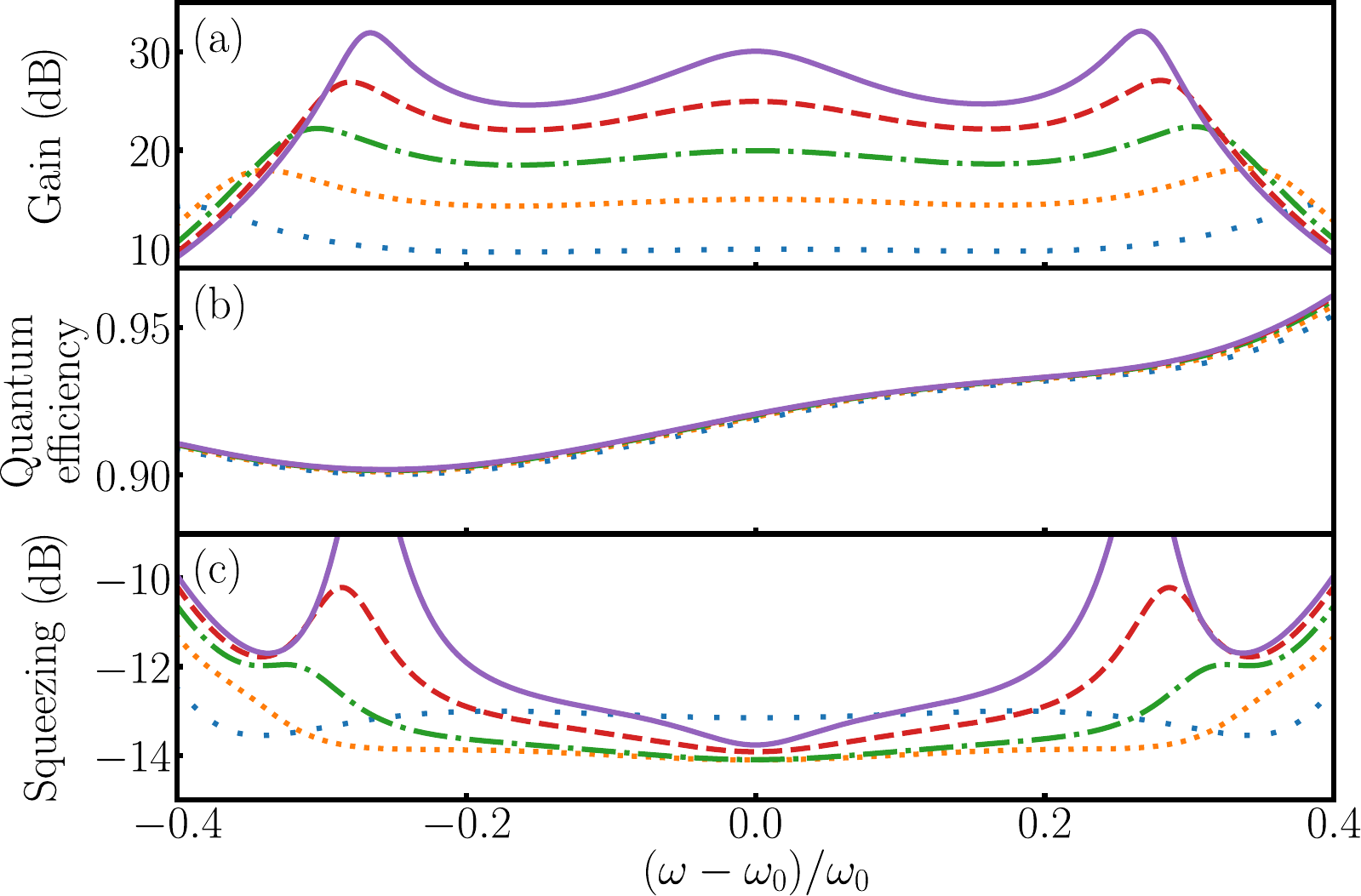}
\caption{Matching of an aluminum SIS junction with $\omega_0/2\pi = 6$ GHz ($\dw\approx 7.2$): phase-preserving gain (a), quantum efficiency (b) and squeezing (c) as a function of the frequency detuning from $\omega_0$ for the dc voltage $eV_\text{dc} = 2\Delta-1.999\hbar\omega_0$ and five different values of ac-voltage amplitude, $\rho = eV_\text{ac}/2\hbar \omega_0 \approx 0.125$ (dotted line), $0.144$ (long-dotted line), $0.155$ (dashed-dotted line), $0.162$ (dashed line), $0.166$ (solid line). These values of the ac-voltage amplitude are chosen such that the phase-preserving gain at $\omega=\omega_0$ is equal to, respectively, $10,15,20,25,30$~dB. The values of the matching-circuit parameters are: $C_1^\prime=0$, $L_1^\prime\approx3.39$~nH, $C_2\approx424$~fF, $L_2\approx1.87$~nH, $C_1=C_3\approx472$~fF and $L_1=L_3\approx1.68$~nH. \label{fig7}}
\end{center}
\end{figure}

Figure~\ref{fig7} shows the (a) phase-preserving gain, (b) quantum efficiency and (c) squeezing for various values of $eV_\text{ac}/2\hbar\omega_0$ at $eV_\text{dc} = 2\Delta - 1.999\hbar\omega_0$.   As shown in Fig.~\ref{fig7}(a), the matching scheme dramatically increases the 3 dB bandwidth of the phase-preserving gain to 4.3~GHz, for a gain of $20$~dB, from the $1.2$~GHz bandwidth obtained without the matching circuit [see Fig.~\ref{fig4}(a)]. The gain ripples are induced by the Chebyshev ripples of the filter reflection coefficient; the higher the gain the higher the amplitude of the ripples. The two extra peaks at the band edges are due to the diminution of $\re Y_M$ at the band edges, which at some frequency cancel the real part of the denominator of Eq.~\eqref{gain-Y-matching}. Reduction of the band edge peaks can be done by adding an imaginary part to the denominator of Eq.~\eqref{gain-Y-matching} at the band edges~\cite{matthaei-paramp}. Here, this is achieved by slightly overestimating the first inductance $L_1^\prime$. 

The compensation of the frequency dependence of $\im Y_0$ has not only increased the 3 dB bandwidth, but also enhanced quantum efficiency [Fig.~\ref{fig7}(b)] and maximum squeezing [Fig.~\ref{fig7}(c)]. The improvement of the quantum efficiency and squeezing is due to the reduction of ac-voltage amplitude and, consequently, single-photon dissipation $\re Y_0$. For instance, a gain of 20~dB is now achieved with an ac-amplitude of $eV_\text{ac}/2\hbar\omega_0=0.155$ (dashed-dotted line), a reduction of 13\% in comparison to the amplifier without the matching circuit [Fig.~\ref{fig4}(a)]. Squeezing now has complex shape at higher gains but remains extremely flat for $G\leq15$~dB.

It is remarkable to obtain such an improvement in a wide bandwidth by using a simple 3-poles filter, designed to match purely capacitive or inductive elements of passive circuits. Indeed, the SIS parametric amplifier is an active circuit with a non-linear dynamical susceptance $\im Y_0$. Such filters can be fabricated directly on-chip either by realizing lumped capacitors and inductors, or using distributed elements such as $\lambda/4$ lines.

\subsection{Dynamical range \label{dynrange}}

An important characteristic of any parametric amplifier is its dynamical range, which determines how many photons can be amplified without saturating the device. In general, the dynamical range is limited by higher order non-linearities of the Hamiltonian and depletion of the pump \cite{roy-devoret-prb2018}. Thus, similarly to SIS mixers, we expect that the dynamical range for a single-junction amplifier to be small. However, we can ensure that amplification of vacuum fluctuations will not saturate the SIS amplifier. This requires the power of amplified vacuum fluctuations, $P_\text{vac}\approx GB\hbar\omega_0$, to be lower than the power delivered by the pump, $P_\text{pump} \approx V_\text{ac}^2\re Y_0(2\omega_0)/2$. However, for amplification in a wide bandwidth, as shown in Fig.~\ref{fig7}, we obtain a ratio $P_\text{vac}/P_\text{pump}\approx1$. This can be overcome by use of a standard approach to design SIS mixers: using $N$ identical SIS junctions in series. It is shown in Ref.~\onlinecite{trucker-feldman-rmp1985} that such an array is equivalent to a single junction with a pump and saturation power $N^2$ times larger. The underlying mechanism is easily understood: the impedance of a series array of $N$ SIS junctions is simply $N$ times the impedance of a single junction. Thus, adding $N$ junctions of resistance $R_T/N$ in series, the admittance remains the same. However, as the photon-assisted effects, necessary for parametric amplification and squeezing, are proportional to $eV_\text{ac}/2\hbar \omega_0$, with $V_\text{ac}$ the applied pump voltage to each junction, the \emph{total} pump voltage across the array has to be $N$ times large. As a consequence, we expect an array of only 5 to 10 SIS junctions to give an experimental relevant dynamical range.

One alternative to increase the dynamical range of a single-junction amplifier is to use a very low-impedance junction ($ <5\Omega$), together with an impedance transformer to match the low-impedance junction with the $50\Omega$ transmission line. Indeed, decreasing the junction impedance increases the pump power $P_\text{pump}$ by the same amount. In practice, a broadband impedance transformation can be implemented either by an extra network of quarter wavelength transmission lines or directly into the filter matching network by using shunt resonators and admittance inverters in between~\cite{matthaei-book}.

\section{Final remarks \label{Final remarks}}

We have proposed a near-quantum-limited broadband amplifier and squeezer based on the photon-assisted tunneling of quasiparticles in a SIS junction. This device can function as a phase-sensitive or phase-preserving amplifier. The gain can be tuned by ac-voltages amplitude to reach gain-bandwidth products of approximatally 12 GHz in the 20-30~dB range, which is 87\% larger than the impedance-engineered Josephson parametric amplifier \cite{roy-vijay-apl2015}. This device is also a source of far separated two-mode squeezing with 3 dB bandwidth of approximately 5 GHz and $-13$ dB of squeezing at the center frequency. Moreover, gain and two-mode squeezing can be fine-tuned in-situ by simply changing the pumping tone amplitude and frequency. For few GHz bandwidth applications, a matching impedance circuit was developed. The proposed matching scheme allows for 3 dB bandwidth of 4~GHz. Also, we estimate that dynamical range of the such a broadband amplifier can be enhanced by replacing the single junction by an array of 5 to 10 SIS junctions. To conclude, the design and fabrication simplicity of this SIS amplifier, together with its operational-mode flexibility, makes it a versatile near-quantum-limited microwave amplifier and squeezer, that can be easily integrated in many quantum microwave experiments.

\section*{Acknowledgments}

U.C.M thanks S. Boutin, A. L. Grimsmo, and M. Westig for fruitful discussions, and the Quantronics group for hospitality. U.C.M., S.J., B.R. and A.B. were supported by the Canada First Research Excellence Fund and NSERC. S.J. and B.R. were supported by Canada Excellence Research Chairs, the Government of Canada, Qu\'ebec MEIE, Qu\'ebec FRQNT via INTRIQ, Universit\'e de Sherbrooke via EPIQ, and the Canada Foundation for Innovation. The research at CEA Saclay received funding from the European Research Council under the European Union’s Horizon 2020 program (European Research Council Grant Agreement No. 639039) and support from the ANR AnPhoTeQ research contract.

%

\end{document}